\documentclass{emulateapj}
\usepackage{placeins}
\usepackage{graphicx}

\def\ba{\begin{eqnarray}}
\def\ea{\end{eqnarray}}

\shorttitle{planet-disk}

\begin{document}

\title{Planet-Disk interaction in 3D: the importance of buoyancy waves}

\author{Zhaohuan Zhu \altaffilmark{1},
James M. Stone \altaffilmark{1}, and Roman R. Rafikov \altaffilmark{1}}

\altaffiltext{1}{Department of Astrophysical Sciences, Princeton University, Princeton, NJ, 08544}
\email{zhzhu@astro.princeton.edu, jstone@astro.princeton.edu, rrr@astro.princeton.edu}

\newcommand\msun{\rm M_{\odot}}
\newcommand\lsun{\rm L_{\odot}}
\newcommand\msunyr{\rm M_{\odot}\,yr^{-1}}
\newcommand\be{\begin{equation}}
\newcommand\en{\end{equation}}
\newcommand\cm{\rm cm}
\newcommand\kms{\rm{\, km \, s^{-1}}}
\newcommand\K{\rm K}
\newcommand\etal{{\rm et al}.\ }
\newcommand\sd{\partial}
\newcommand\mdot{\rm \dot{M}}
\newcommand\rsun{\rm R_{\odot}}
\newcommand\yr{\rm yr}

\begin{abstract}
We carry out local three dimensional (3D) hydrodynamic simulations of  
planet-disk interaction in stratified disks with varied thermodynamic 
properties. We find that whenever the Brunt-V\"ais\"al\"a 
frequency ($N$) in the 
disk is nonzero, the planet exerts a strong torque on the disk in the
vicinity of the planet, with a reduction in the traditional``torque
cutoff''.  In particular, this is
true for adiabatic perturbations in disks with isothermal density
structure, as should be typical for centrally irradiated protoplanetary
disks. We identify this torque with buoyancy waves, which are excited
(when $N$ is non-zero) close to the planet, within one disk scale
height from its orbit.  These waves give rise to density
perturbations with a characteristic 3D spatial pattern which is in close
agreement with the 
linear dispersion relation for buoyancy waves.
The torque due to these waves can amount to as much as several
tens of per cent of the total planetary torque, which is not expected
based on analytical calculations limited to axisymmetric or low-$m$
modes. Buoyancy waves should be ubiquitous around planets in the
inner, dense regions of protoplanetary disks, where they might
possibly affect planet migration.  \end{abstract}



\section{Introduction}
\label{sect:intro}

The gravitational potential of a planet surrounded by a protoplanetary 
disk is known to give rise to non-axisymmetric density waves, which
propagate away from the planet carrying angular momentum. 
Gravitational coupling with these density perturbations
exerts a torque on the planet, which
might lead to its migration if the one-sided torques produced by the 
inner and outer disks do not cancel. Pioneering linear calculations 
by \cite{GT} have shown that, neglecting this possible small 
asymmetry, the one-sided torque acting on a planet moving on a circular 
orbit in a razor-thin (2D) disk is 
\begin{equation}
T= q_{\rm 2D} (GM_{p})^{2}\frac{\Sigma_{0}R_{p}\Omega_{p}}{c_{s}^{3}},
~~~q_{\rm 2D}\approx0.93,
\label{eq:eqt}
\end{equation}
where $M_p$, $R_{p}$ and $\Omega_p $ are the planetary mass, semi-major 
axis and angular frequency respectively, 
$c_s\equiv (\partial P/\partial\rho|_S)^{1/2}$ is the adiabatic sound 
speed, and $\Sigma_{0}$ is the unperturbed disk surface density.
Most of this torque is excited at Lindblad resonances corresponding to 
the $m\sim R_{p}/h\gg 1$ azimuthal harmonic of the planetary potential and 
located at a distance $\sim h=c_{s}/\Omega_{p}$ away from the 
planetary orbit. Closer to the planet, at separations $\lesssim h$ the
excitation torque density decreases exponentially --- the so-called 
``torque cut off'' phenomenon.

In three-dimensional (3D) stratified disks a variety of waves can be 
excited (Lubow \& Pringle 1993, Korycansky \& Pringle 1995, Lubow \& Ogilvie 1998), 
including buoyancy waves whenever 
the Brunt-V\"ais\"al\"a frequency
\ba
N\equiv \left[\frac{g}{\gamma}\frac{\partial}{\partial z}{\rm ln}
\left(\frac{P_{0}}{\rho_{0}^{\gamma}}\right)\right]^{1/2}
\label{eq:N2}
\ea
is non-zero. Global analytical calculation of the wave properties for all modes
are available only for one special case, in which both the vertical disk 
structure and the equation of state (EOS) are isothermal, and $N=0$. 
In this case,
the result (\ref{eq:eqt}) still holds but with $q_{\rm 2D}$ 
lowered to $q_{\rm 3D}\approx 0.25 q_{\rm 2D}=0.37$
as a result of averaging the planetary potential 
over the vertical disk structure \citep{TM}. For other combinations 
of the vertical disk structure and EOS, in particular those characterized 
by non-zero $N$, analytical results are limited only to either axisymmetric or low-$m$ 
modes excited far from the planet. In that case \cite{LO} found that although
buoyancy waves are responsible for a fraction of the total 
torque excited by the planet, the total torque from all modes is the
same as that given by traditional Lindblad torque formulae.

In this paper we carry out 3D simulations of stratified disks 
characterized by non-zero $N$ with embedded low-mass planets  
to explore global excitation of buoyancy waves and their role 
in angular momentum exchange with the planet.


\section{Numerical Method}
\label{sect:numerics}

The numerical tool we used is Athena \citep{SGTHS}, a grid-based 
code with a higher-order Godunov scheme, piecewise parabolic method 
(PPM) for spatial reconstruction, and the corner transport upwind (CTU)
method for multidimensional integration. We 
use the stratified shearing-box set-up implemented in Athena 
\citep{SG}. Since the potential vorticity is zero in the shearing-box,
the traditional corotation torque is zero (Goldreich 
\& Tremaine, 1979).

A planet is placed at the origin of a Cartesian box 
in $x,y,z$ coordinates. Its smoothed potential is approximated as
\ba
\Phi_{p}(d)=-GM_{p}\frac{d^2+3r_{s}^{2}/2}{(d^2+r_{s}^{2})^{3/2}},
\label{eq:fourthp}
\ea
where $r_{s}$ is the smoothing length and $d=\sqrt{x^2+y^2+z^{2}}$.
This potential converges to the point mass potential as 
$(r_{s}/d)^{4}$ for $d\gg r_{s}$. 
In most runs $r_{s}=0.125 h$, which
is resolved by 4 grids cells with our normal resolution of 32/$h$.

\begin{figure}
\includegraphics[width=0.9\textwidth]{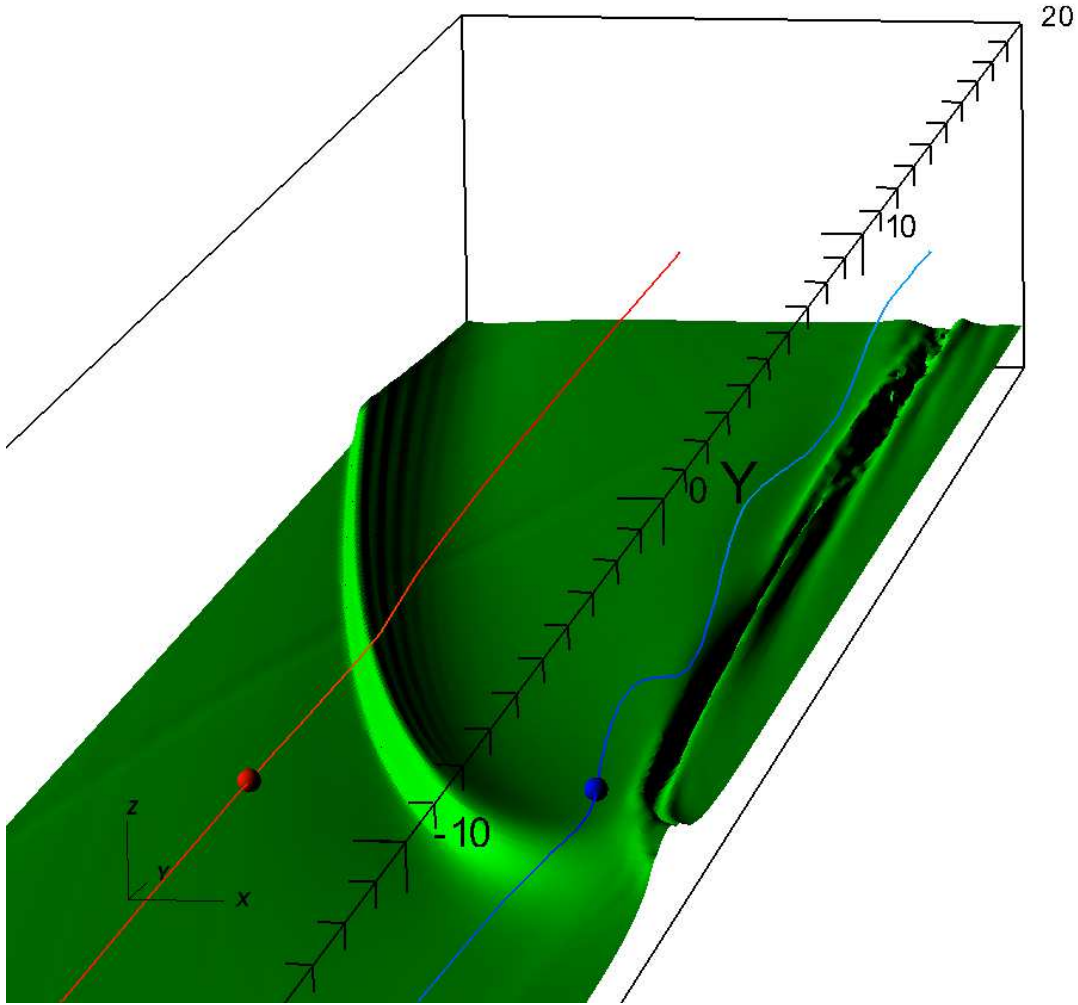} \\
\caption{The iso-density contour of $\rho=0.8\rho_{00}$ for an IA 
disk model. Planetary mass is increased to $M_{p}=0.3 M_{th}$ to 
visually enhance the effect of buoyancy waves. Buoyancy waves cause 
ray-like density disturbances close to $x=0$ (right side 
of the box), which are also shown in Fig. \ref{fig:fig6}. Two 
streamlines passing through (-0.5, 0, 1)(blue) and (-3, 0, 1)(red) demonstrate 
vertical oscillation due to buoyancy waves and the small velocity perturbation of sound waves.}
\label{fig:fig1}
\end{figure}

To reduce the computation time we take advantage of the symmetry of 
the problem and only simulate the disk for $x=[-X,0]$, $y=[-Y,Y]$, 
and $z=[0,Z]$, see Figure \ref{fig:fig1}. We take $X=5h$, $Y=20h$ 
so that the planetary wake located at $y\sim-3x^{2}/4h$ 
fits well inside the box. We vary $Z$ in different runs (see Table \ref{tab2}) 
in such a way that the density at $z=Z$ is $\sim$5 orders of magnitude 
lower than the midplane density.

We use different boundary conditions (BCs) at 
different box faces: outflow BC at the $y=Y$ face, and ``fixed state'' 
BC (meaning that all physical variables in the ghost zones are fixed at 
their unperturbed Keplerian values) at  $x=-X$ and $y=-Y$ faces 
(for one model we also use periodic BC in $y$-direction, see 
Table \ref{tab2}).
At the $x=0$ face we employ a ``symmetric'' BC: the $x=0$ face has 
been divided into two portions --- $y>0$ and $y<0$, and the variables 
in the last 4 active zones of one portion are copied into the ghost 
zones in the other portion symmetric with respect to $x=y=0$; 
velocities (or momenta) in $x$ and $y$ directions are copied with 
the opposite sign. We use a reflecting BC at $z=0$ since both 
the planetary potential and the disk structure are symmetric 
with respect to $z$. At $z=Z$ we use an outflow BC but the 
densities are extrapolated assuming hydrostatic equilibrium 
into the ghost zones, and to prevent inflow 
if $v_{z}<0$, then we set $v_{z}$ to 0.


\subsection{Physical setup}
\label{sect:phys_setup}

In our calculations we use both an isothermal ($\gamma$=1 in 
$p\propto \rho^{\gamma}$), 
and an adiabatic EOS ($\gamma=5/3$ or $7/5$). The unit of length 
in our simulations is defined via the adiabatic scale 
height $h=c_s/\Omega_p=\sqrt{\gamma}c_{iso}/\Omega$, where
$c_{iso}\equiv (kT/\mu)^{1/2}$ and $c_s$ is the adiabatic sound speed 
calculated using the 
midplane temperature.  Obviously, $c_s=c_{iso}$ for an isothermal
EOS ($\gamma=1$). This choice keeps sound waves traveling 
the same distance in a given amount of time in different
models so that the wake position is always the same.

The vertical structure of a protoplanetary disk is typically 
determined by irradiation from 
the central star so that is vertically isothermal 
\ba
\frac{\rho_{0}(z)}{\rho_{00}}={\rm exp}
\left(-\frac{z^{2}\Omega^{2}}{2c_{iso}^{2}}\right),
\label{eq:iso}
\ea
where $\rho_{00}$ is the midplane density. The value of $c_{iso}$ 
is determined by the central irradiation flux \citep{KH,CG}. 

We also consider thermally stratified disks in which the disk midplane
is hotter than its atmosphere, which is expected if viscous heating 
is important. Including the fact that such disks will eventually 
become isothermal at hight $z$,
 the vertical structure of the disk is better 
represented as
\ba
p_{0}(z)=c_{s,s}^{2}\rho_{0}(z)\left[\left(\frac{\Gamma-1}
{\Gamma}\right)\left(\frac{\rho_{0}(z)}{\rho_{s}}
\right)^{\Gamma-1}+1\right],
\label{eq:pd}
\ea
where $c_{s,s}$ and $\rho_{s}$ ( set as 0.01 $\rho_{c}$) represent  the sound speed and density 
at the transition where the disk becomes isothermal
 \citep{LPS, BOLP}. 
However, unlike \cite{BOLP}, we do not necessarily set $\Gamma$ 
in Eq. (\ref{eq:pd}) the same as $\gamma$ in the EOS. We derive 
$\rho_{0}(z)$ by solving the equation of vertical 
hydrostatic equilibrium with Eq. (\ref{eq:pd}) using the 
Newton-Raphson method. When $\Gamma\to 1$ the isothermal disk 
structure is recovered. 

We run a set of five models up to 10 orbits with different disk structure and 
EOS, summarized below and in Table \ref{tab2}.

(a) Isothermal disk structure (\ref{eq:iso}) with isothermal 
EOS (model II). 

(b) Disk structure (\ref{eq:pd}) with adiabatic EOS and 
$\Gamma=\gamma=5/3$ (model AA). 

(c) Isothermal disk structure with adiabatic EOS having 
$\gamma=5/3$ (model IA). This setup is typical for a centrally
irradiated disk in which density perturbations are adiabatic.

(d) Analogous to (4) but with $\gamma=7/5$ (model IA2), designed 
to illustrate the effect of adiabatic index on the buoyancy wave 
coupling.
 
(e) Disk structure (\ref{eq:pd}) with $\Gamma=5/3$ and 
adiabatic EOS with $\gamma=7/5$ (model PA), to illustrate 
viscously heated accretion disks. 

In the first two models $N=0$ so that buoyancy waves are absent,
while the last three models feature non-zero $N$ and support
buoyancy waves. All models were run at a resolution of 32/$h$, but 
we also run model (3) at higher resolution (64/$h$) to verify 
numerical convergence. 

The planet mass is $M_p=0.0058 M_{th}$, where $M_{th}$ is the 
thermal mass
\begin{equation}
M_{th}\equiv\frac{c_{s}^{3}}{G\Omega_{p}}.
\end{equation}
Thus, disk-planet coupling is well inside the linear regime. Density 
wave generated by such a planet in a 2D disk would shock at 
$|x|=6h$ \citep{GR}, which is outside our box.


\section{Results}
\label{sect:results}


\subsection{Without Buoyancy Waves}

First we show the results for models with no buoyancy waves (II and AA).
The spatial distributions of the excitation torque density $dT/dx$ 
(the amount of torque excited per unit radial distance) for these 
two models are shown in Fig. \ref{fig:fig8}(a)(b)
as green and orange curves. One can see that $dT/dx$ rapidly goes 
to zero for $|x|\lesssim h$, clearly exhibiting the 
``torque cut-off'' phenomenon \citep{GT} in both models. 

For model II, our simulation agrees with the semi-analytical 
calculation by Takeuchi \& Miyama (1998) remarkably well, and we 
confirm that Eq. (\ref{eq:eqt}) holds in this case with 
$q_{\rm 3D}(II)=0.25  q_{\rm {2D}}$. Although wave structures are 
different in model AA (Lubow \& Pringle 1993), the distribution of 
$dT/dx$ for this model is quite similar in shape to that of 
model II, but the amplitude is slightly different.   


\subsection{With Buoyancy waves}

\begin{figure}
\includegraphics[width=0.9\textwidth]{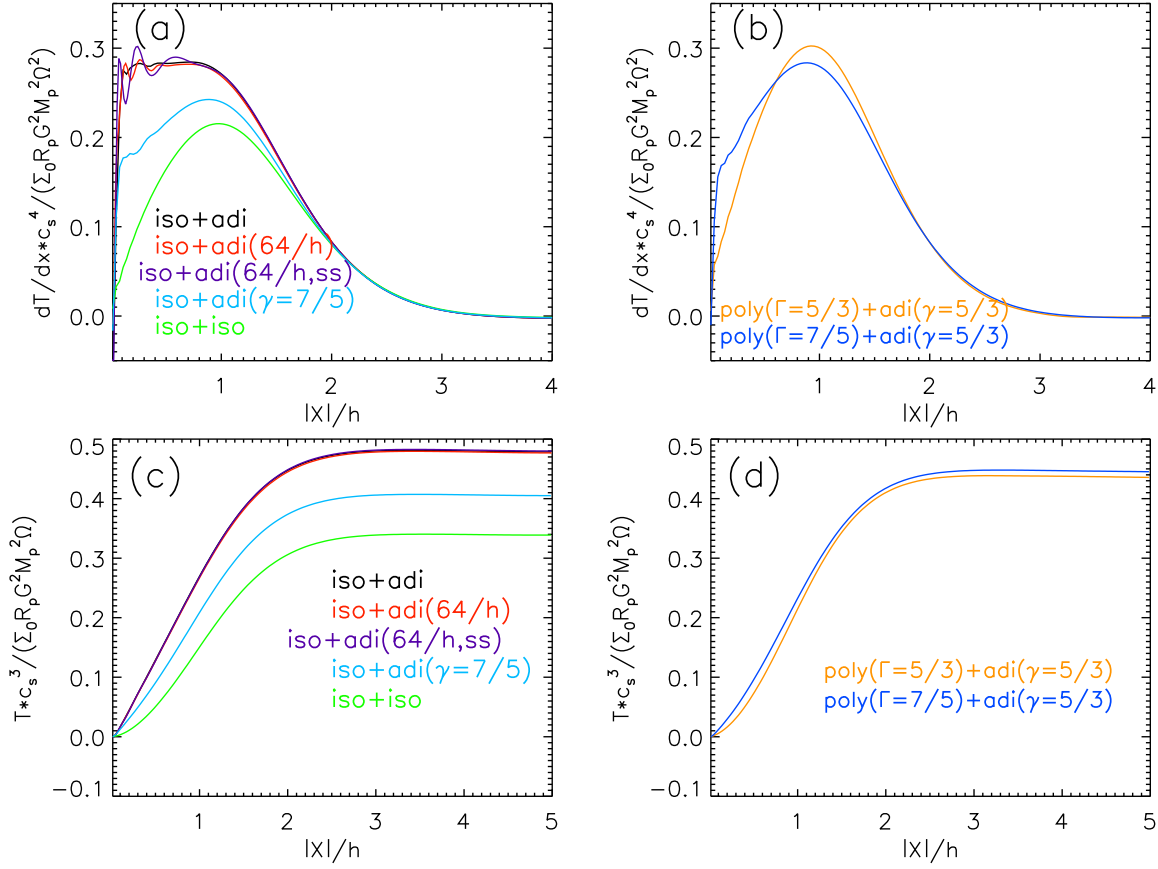} \\
\caption{Torque densities ($dT/dx$, upper panels) and integrated torques (lower panels)
 for disks with isothermal disk structure (left panels) and polytropic disk 
 structure (right panels) with different EOS. 
Models II (green curves) and AA (orange curves) not supporting 
buoyancy waves exhibit a clear torque cut-off at $|x|\lesssim h$. 
All other models support buoyancy waves and show significant 
buoyancy torque near the planet.
}
\label{fig:fig8}
\end{figure}

We next look at the behavior of $dT/dx$ in models with non-zero $N$.
The most important feature of these models, evident in Figure  
\ref{fig:fig8}, is the weak torque cutoff near the planet ---
all show a significant torque contribution near $x=0$, 
including model IA2, which provides the best description for an 
irradiated protoplanetary disk. Since higher-m modes
are excited at Lindblad resonances closer to the planet, the strong torque
close to the planet also suggests the traditional torque cutoff in
Fourier space (Ward 1997) is weaker
when buoyancy waves are present.

The one-sided torque $T$  
(Fig. \ref{fig:fig8}(c)(d)) in models IA and IA2 
is characterized by $q_{\rm 3D}(IA)=0.48$ and 
$q_{\rm 3D}(IA2)=0.41$ respectively, which should be compared with
$q_{3D}(II)=0.34$ for model II having the same vertical structure 
as IA and IA2 but different EOS.
 
\begin{figure}
\epsscale{1.0} \plotone{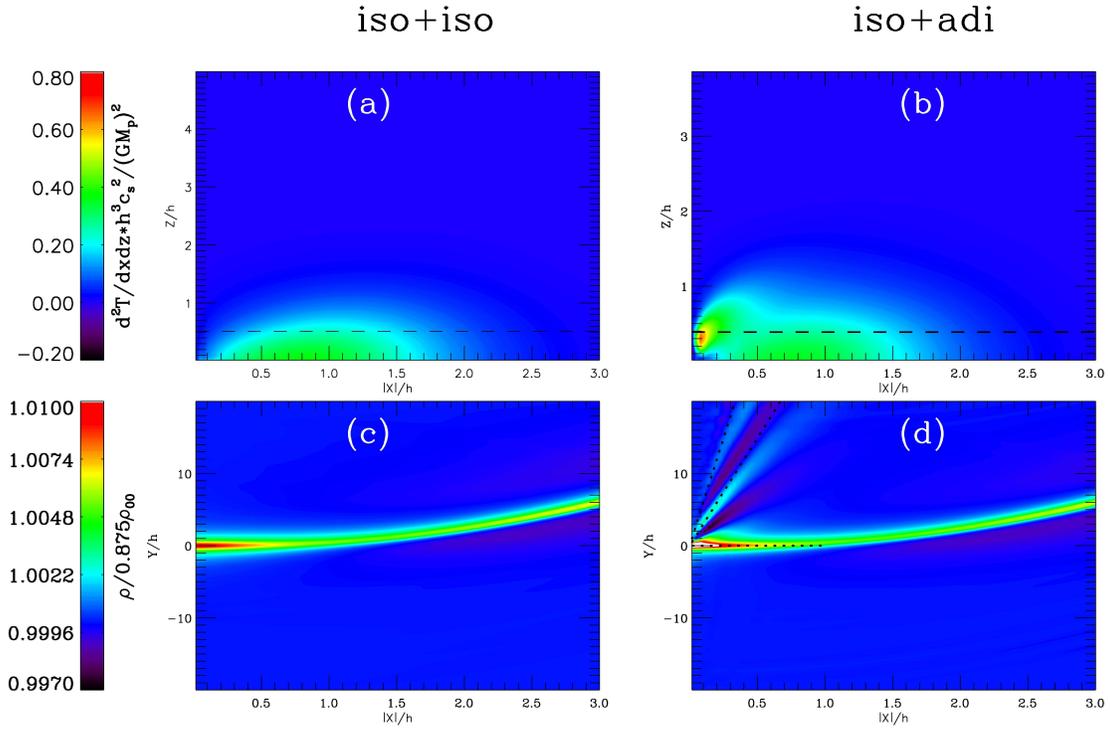} \caption{Upper panels: 
(volume) density of the planet-induced torque integrated over 
$y$-coordinate for models II (left panel) and IA (right panel). 
Strong excess torque due to buoyancy waves close to the planet 
is clearly visible at $z\sim h$ in model IA, which features 
non-zero Brunt-V\"ais\"l\"a frequency. 
Lower panels: density contours in the $xy$ plane 
at $z$ where $\rho_{0}$ equal to 0.8755 $\rho_{00}$ (indicated by the 
dashed line at $z\sim0.5 h$ in (a) and $z\sim 0.4 h$ in (b), h are different in these two cases). 
Geometric locations corresponding to the resonance condition 
for buoyancy waves 
(\ref{eq:resonance}) are shown by dotted lines in model IA, 
and agree quite well with the pattern of density fluctuations 
derived from simulations.
} \label{fig:fig6}
\end{figure}

In order to identify the origin of this excess torque near the 
planet, we integrate the volume density of the torque along $y$ 
direction 
\begin{equation}
\frac{d^{2}T}{dxdz}=\int_{-\infty}^{\infty}dy\delta\rho(x,y,z)\frac{\partial \phi}{\partial y}.
\end{equation}
 This quantity is shown in Fig. \ref{fig:fig6}(a)
for models II and IA. We see that the excess torque in model IA 
comes from the region $z\sim$ 0.4 $h$ and $|x| < h$. In Fig. \ref{fig:fig6}(d)
 we plot the density contours in the $xy$ 
plane at $z=0.4 h$. In clear contrast with model II (Fig. \ref{fig:fig6}(c)), the density 
distribution in model IA exhibits density fluctuations/ridges
{\it close} to the planet which are different from the usual wake 
structure. They extend to $y>0$ and look like rays emanating from 
the origin. It is these density fluctuations 
that contribute to the excess torque. 

\begin{figure}
\epsscale{1.0} \plotone{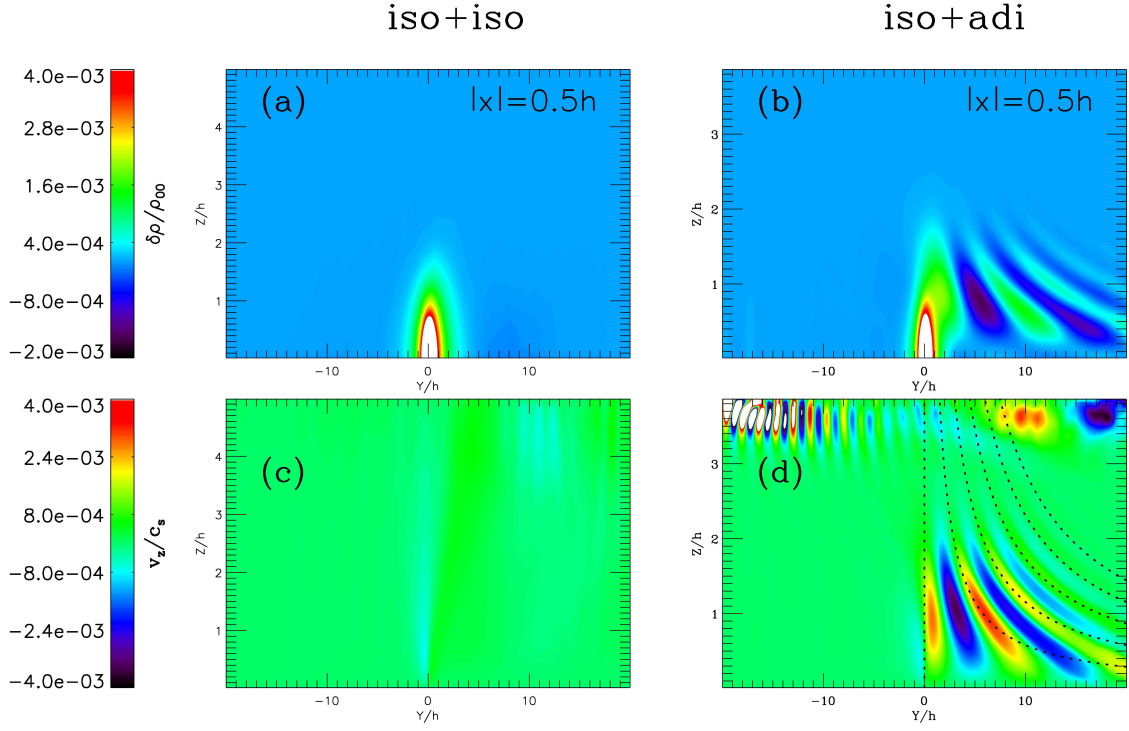} \caption{Upper panels: density 
fluctuations in the $yz$ plane at $|x|=0.5 h$ for models II (left panel) 
and IA (right panel). Lower panels: vertical velocity $v_{z}$ at 
$|x|=0.5 h$ for models II and IA. The buoyancy resonance positions 
(\ref{eq:resonance}) are again plotted as dotted curves. The 
fluctuations of $v_z$ at $z>3h$ are due to the lack of exact hydrostatic equilibrium at  
the upper boundary.
} \label{fig:fig7}
\end{figure}

These fluctuations also  have vertical structure, as shown in 
Fig. \ref{fig:fig7} where we plot $\delta \rho=\rho-\rho_{0}$ 
and $v_{z}$ in the $yz$ plane at $x=0.5 h$ for both models II and IA.  
Again, significant density fluctuations close to the planet 
exist only in case IA. 

Since the only difference between case IA and case II is that 
the case IA has non-zero Brunt-V\"ais\"al\"a frequency $N$,
it is natural to relate these density fluctuations to buoyancy 
waves. To confirm this hypothesis, we note that a mode with 
wavenumber $k_y$ has the buoyancy frequency $N$
whenever the condition 
\begin{equation}
2Ak_{y}x=N
\label{eq:N1}
\end{equation}
is fulfilled. For comparison, the Lindblad resonance condition is 
$2Ak_{y}x=\pm \kappa$. In the disk with isothermal structure and 
adiabatic EOS (models IA and IA2), we have
\begin{equation}
N=\frac{\Omega z}{h}\sqrt{\frac{\gamma-1}{\gamma}}.
\label{eq:N3}
\end{equation}
With $\gamma=5/3$, and $\lambda_{y}=2\pi/k_{y}$,  Eqs. \ref{eq:N1} 
and \ref{eq:N3} can be combined to derive
\begin{equation}
\frac{\lambda_{y}}{h}=11.543 \frac{x}{z}.\label{eq:lambday}
\end{equation}
If we assume the phase of buoyancy waves is 0 at $x$=0, 
the geometric location of the constant phase $2n\pi$ ($n$ 
is integer) is given by 
\begin{equation}
\frac{y}{h}=11.543 \frac{x}{z}n\,\,\,\,{\rm and}\,\,\, n=1,2,3,...\,.
\label{eq:resonance}
\end{equation}
Curves corresponding to Eq. \ref{eq:resonance} are drawn in the 
lower right panels of Figs. 
\ref{fig:fig6} and \ref{fig:fig7}. They agree
with the pattern of the density perturbation and $v_{z}$ quite 
well, strongly suggesting that these fluctuations
and the excess torque are related to buoyancy waves (their dissipation 
or excitation). 

The torque density and integrated torque for model IA2 
(isothermal disk with adiabatic EOS and $\gamma=1.4$)
plotted as the cyan curves in Fig. \ref{fig:fig8} also exhibit
the excess torque density due to buoyancy waves close to the planet.
However, the integrated torque in this case is characterized 
by $q_{3D}(IA2)\approx 0.41$,
which is lower than $q_{3D}(IA)$, and the buoyancy wave 
coupling to the planetary potential is weaker for smaller $\gamma$ as
seen from $dT/dx$ curve. 

Blue curves in the right panels of Fig. \ref{fig:fig8} show
$dT/dx$ and $T(x)$ for model PA. There is again an apparent torque
excess near the planet, which is not surprising since $N\neq 0$ 
in this model allowing buoyancy waves to be excited. 


\section{Discussion}
\label{sect:disc}
The phenomenon of torque cut-off in 2D disks is due to the 
fact that high-$k_y$ modes 
($k_y\gtrsim h^{-1}$, excited close to the planet) 
of rotation-modified sound
waves couple poorly to the planetary 
potential (even though it is strongest there).
 However, the dispersion relation for 
buoyancy waves is very different from that of the waves in 2D
disks. Based on Eq. (\ref{eq:lambday}), the pattern in our simulations should have 
$k_{y}\approx h^{-1}$ at $x\sim0.05 h$ and $z=0.1 h$, resulting in strong 
coupling between buoyancy waves and the planetary 
potential. This may explain the lack of torque cutoff in models
with non-zero $N$. A further investigation will be presented in Zhu \etal (in prep.).

Since the buoyancy torque is present very close to the planet, 
its strength could be sensitive 
to the potential softening length $r_s$ in the simulation. Thus, we re-ran model 
IA with a smaller smoothing length ($r_{s}=1/16 h$) but obtained essentially 
the same result (purple curves in Fig. \ref{fig:fig8}), 
suggesting that our numerical results are robust.


\subsection{Comparison with previous studies}

Previous analytical and semi-analytical studies of buoyancy waves 
(Lubow \& Pringle 1993, Korycansky \& Pringle 1995,  Lubow \& Ogilvie  
1998) have been limited to axisymmetric waves or low-$m$ modes 
excited far away from the planet, at $|x|\gg h$. These studies 
have demonstrated such modes to play rather insignificant role in 
planet-disk angular momentum exchange. 

The main result of this work is that coupling of the 
planetary
potential to the {\it locally-excited} ($|x|\lesssim h$) buoyancy 
waves channels a considerable amount of angular momentum into these 
modes. As a result, the buoyancy torque strongly contributes to 
the planetary torque, at the level of tens of per cent. This result is different from 
analytical expectations extrapolated from low-$m$ buoyancy modes \citep{LO}.

Most previous 3D simulations of planet-disk interaction 
explored isothermal disk structure with an isothermal EOS 
(e.g. Bate \etal 2003, D'Angelo \& Lubow 2010).
An adiabatic EOS has been explored in \cite{PM}. However, because
of the global geometry used in their work, the buoyancy torque may have 
been hard to distinguish from the corotation torque, see 
\S \ref{sect:migration}.


\subsection{Existence and Signatures of Buoyancy Waves}

The vertical structure of protoplanetary disks around T Tauri
stars should be close to isothermal given the dominant
role of stellar irradiation in their thermal balance. 
Existence of buoyancy waves in such disks depends on the
thermodynamics of the fluid perturbations on timescales 
$\sim N^{-1}$. Density perturbations induced by planets
result in temperature fluctuations which tend to be
erased by radiative diffusion. If the cooling time $t_c$ of
such perturbations is longer than $N^{-1}$ then modes are 
adiabatic with $\gamma>1$ (as in models IA and IA2) and 
buoyancy waves should be present. In the opposite case of
$t_c N\lesssim 1$ fluid perturbations behave as isothermal
($\gamma\approx 1$) and buoyancy waves are not excited, as 
in model II. 

We expect that the separation between the two regimes
should occur between several to several tens of AU,
depending on the disk properties, and that buoyancy waves 
should be efficiently excited by relatively close-in planets. 
Further out $t_c N\lesssim 1$ 
and excitation of buoyancy waves should be suppressed. Note,
that both $t_c$ and $N$ are in general functions 
of $z$, see equation (\ref{eq:N3}). 
This makes cooling considerations dependent not only 
on the radius but also on the height in the disk.  

Whenever buoyancy waves are efficiently excited by planets 
they can reveal themselves via associated vertical motions
(see Fig. \ref{fig:fig1}) which should give rise to corrugations 
of the disk surface aligned with density ridges seen in 
Fig. \ref{fig:fig6}. Near-IR imaging of stellar light scattered 
by such perturbations at very high resolution (on scales $\sim 
h$) can reveal  even low-mass planets embedded in 
protoplanetary disks.


\subsection{Migration}
\label{sect:migration}

The rate at which a planet migrates is determined by the 
imbalance of one-sided torques exerted on it by the inner 
and outer parts of the disk, caused by the {\it global} 
radial gradients of the disk density and temperature 
(Ward 1986). The net torque can also be caused by the 
{\it local} asymmetries in the disk properties caused
by the presence of the planet itself, such as the 
temperature perturbations due to shadowing near the planet 
(Jang-Condell \& Sasselov 2005), or
the planet's own heat output caused by e.g. the 
accretion of planetesimals. Such asymmetries generated 
very close to the planet may not 
strongly affect the net standard 
Lindblad torque because of the torque cutoff at such 
separations. However, their effect on the {\it net 
buoyancy torque} excited primarily at small $x$ may 
be disproportionately large and considerably affect the 
planetary migration. 

This statement is only true if the buoyancy waves 
observed in this work can {\it propagate} and deposit 
their angular momentum far from the corotation region. 
Otherwise their angular momentum accumulates in the 
narrow annulus around the planetary orbit and the corresponding
torque on the planet may be subject to saturation, like 
the standard corotation torque (Balmforth \& Korycansky 2001; 
Masset 2001, 2002). This would eliminate the effect of 
the buoyancy waves on the planetary migration. We 
will investigate these possibilities in Zhu \etal (in prep).  

Direct measurement of the buoyancy torque effect on the 
migration speed would require global disk-planet calculations, 
in which corotation torque appears as well. It may then be 
difficult to separate the effects of the buoyancy and 
corotation torques in global simulations. However, the strength 
of the corotation torque depends on the radial gradients 
of specific vortensity and entropy across the 
horseshoe region (Paardekooper \& Mellema 2006; 
Baruteau \& Masset 2008; Paardekooper \& Papaloizou 2008). 
By properly setting the disk properties to nullify these 
gradients one can eliminate the corotation torque in global 
simulations, thus isolating the contribution due to the 
buoyancy torque.


\subsection{Wave dissipation and gap opening}

Gap opening by massive planets depends not only on the 
wave excitation but also on wave {\it damping} as a 
means of transferring angular momentum to the disk 
fluid (Lunine \& Stevenson 1982; 
Rafikov 2002). The global density wake excited 
by planet is thought to 
dissipate primarily via shock damping (Goodman \& Rafikov 2001), but buoyancy
waves are very distinct from rotation-modified 
sound waves and should dissipate differently 
(Lubow \& Ogilvie 1998; Bate \etal 2002). 
Channeling of the wave action 
(Lubow \& Ogilvie 1998) may considerably 
speed up their nonlinear evolution resulting in 
more efficient damping. Since we showed that buoyancy 
waves carry good fraction of the total angular 
momentum flux, understanding their damping
mechanism and spatial pattern of dissipation
may be important for clarifying the issue of gap opening 
by planets (Zhu \etal in prep). 

Our work suggests buoyancy waves can play an important role
in planet migration and gap opening which demands further studies.


\acknowledgements
Authors are indebted to Jeremy Goodman, Steve Lubow,
and Gordon Ogilvie for helpful comments and suggestions. This 
work was supported by NSF grant AST-0908269 and Princeton University.
This research was supported in part by the NSF through TeraGrid
resources provided by the Texas Advanced Computing Center and
the National Institute for Computational Science under grant number
TG-AST090106.


\begin{table}
\begin{center}
\caption{Models \label{tab2}}
\begin{tabular}{cccccccc}

\tableline\tableline
Case &   structure  &  EOS & y-direc.  & Brunt-V\"ais\"al\"a  & Torque & Z domain size & Resolution\\
           &          &        $\gamma$             & boundary & $N$ & coefficient (q)\tablenotemark{a} & & X$\times$Y$\times$Z\\
\tableline
II & isothermal & 1 & in/outflow & 0 & 0.34 & [0, 5 h] &160$\times$1280$\times$160\\
AA & poly. $\Gamma=5/3$ & 5/3 & in/outflow & 0 & 0.44 & [0, 2 h] & 160$\times$1280$\times$64\\
AAper & poly. $\Gamma=5/3$ & 5/3 & periodic & 0 & 0.44 & [0,2 h] & 160$\times$1280$\times$64\\
IA  & isothermal & 5/3 & in/outflow & $\ne$0 & 0.48 & [0, 3.873 h] & 160$\times$1280$\times$160\\
IAper  & isothermal & 5/3 & periodic & $\ne$0 & 0.46 & [0, 3.873h] & 160$\times$1280$\times$160\\
IA (64/h)  & isothermal & 5/3 & in/outflow & $\ne$0 & 0.48 & [0, 3.873h]  & 320$\times$2560$\times$320\\
IAss (64/h, r$_{s}$=1/16 h)  & isothermal & 5/3 & in/outflow & $\ne$0 & 0.48 & [0, 3.873h]  & 320$\times$2560$\times$320\\
IA2  & isothermal & 7/5 & in/outflow & $\ne$0 & 0.41& [0, 3.873h] & 160$\times$1280$\times$160 \\
PA & poly. $\Gamma=7/5$ & 5/3 & in/outflow & $\ne$0 & 0.45  &[0, 2.5h] & 160$\times$1280$\times$80\\
\tableline
\end{tabular}
\end{center}
\tablenotetext{a}{ $q$ as in $T=q(GM_{p})^{2}\Sigma_{0} R_{p} \Omega/ c_{s,adi}^{3}$}
\end{table}

\end{document}